\documentstyle[preprint,eqsecnum,aps,floats,epsfig]{revtex}
\def\Journal#1#2#3#4{{#1} {\bf #2}, #3 (#4)}

\def\NPB{{ Nucl. Phys.} B}

\def\PRL{ Phys. Rev. Lett.}
\def\PRD{{ Phys. Rev.} D}
\def\ZPC{{ Z. Phys.} C}

\begin{document}
\draft
%
%
%
%
\topmargin -0.5in
\makeatletter
\def\maketitle{\par
\begingroup
\let\cite\@bylinecite
\def\thefootnote{\fnsymbol{footnote}}%
\if@twocolumn
\twocolumn[\@maketitle\vskip2pc]%
\else
\newpage
\epsfysize3cm
\epsfbox{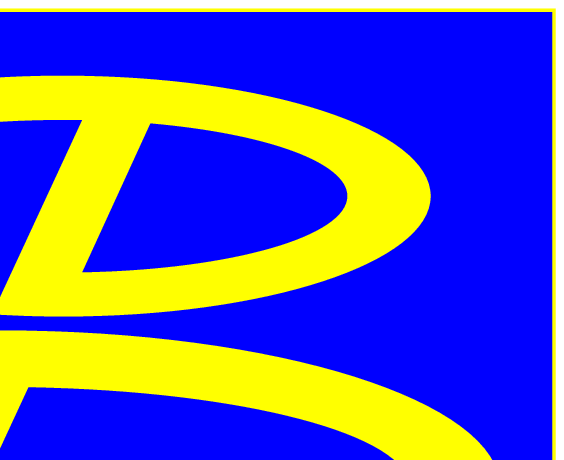}    
\global\@topnum\z@ %
\@maketitle
\fi
\thispagestyle{plain}\@thanks
\endgroup
\def\thefootnote{\arabic{footnote}}%
\setcounter{footnote}{0}%
\let\maketitle\relax \let\@maketitle\relax
\let\@thanks\relax \let\@authoraddress\relax \let\@title\relax
\let\@date\relax \let\thanks\relax
}
\makeatother

\tighten

\draft 


\preprint{\tighten\vbox{\hbox{\hfil KEK preprint 2001-118}
                        \hbox{\hfil Belle preprint 2001-13}
                        \hbox{\hfil DPNU-01-30}
                        \hbox{\hfil }\hbox{\hfil }\hbox{\hfil }
}}

\title{{\large
       Observation of the Decay $B \to K \ell^{+} \ell^{-}$}\footnote{submitted to PRL}\\
}
\tighten

\author{
{\normalsize Belle Collaboration\\}
\ \\
  K.~Abe$^{9}$,               
  K.~Abe$^{38}$,              
  R.~Abe$^{28}$,              
  I.~Adachi$^{9}$,            
  Byoung~Sup~Ahn$^{16}$,      
  H.~Aihara$^{40}$,           
  M.~Akatsu$^{21}$,           
  Y.~Asano$^{45}$,            
  T.~Aso$^{44}$,              
  V.~Aulchenko$^{2}$
% ...
\\
}
\address{
\ \\
% ...
$^{2}${Budker Institute of Nuclear Physics, Novosibirsk}\\
% ...
$^{9}${High Energy Accelerator Research Organization (KEK), Tsukuba}\\
% ...
$^{16}${Korea University, Seoul}\\
% ...
$^{21}${Nagoya University, Nagoya}\\
% ...
$^{28}${Niigata University, Niigata}\\
% ...
$^{38}${Tohoku Gakuin University, Tagajo}\\
% ...
$^{40}${University of Tokyo, Tokyo}\\
% ...
$^{44}${Toyama National College of Maritime Technology, Toyama}\\
$^{45}${University of Tsukuba, Tsukuba}\\
% ...
}

\maketitle
\begin{abstract}
We report a search for the flavor-changing neutral current decay 
$B \to K^{(*)} \ell^{+} \ell^{-}$ using a 29.1~fb${}^{-1}$ 
data sample accumulated at the $\Upsilon(4S)$ resonance with 
the Belle detector at the KEKB $e^{+}e^{-}$ storage ring. 
We observe the decay process $B\to K\ell^+\ell^- ( \ell = e, \mu )$, for the first time,
with a branching fraction of 
${\cal B}(B \to K \ell^{+} \ell^{-}) 
 = (0.75^{+0.25}_{-0.21} \pm 0.09) \times 10^{-6}$.

\end{abstract}


\narrowtext
Flavor-changing neutral current (FCNC) processes are forbidden 
at the tree level in the Standard Model (SM), but are induced 
by loop or box diagrams.
If non-SM particles participate in the loop or box diagrams,
their amplitudes may interfere with the SM amplitudes. 
This makes FCNC processes an ideal place to search for new physics. 

The $b \to s$ transition is a penguin-diagram mediated FCNC process.
The CLEO group reported the first observation of the
$B \to X_s\gamma$ radiative penguin decay \cite{CLEO-radiative}.
The measured branching fraction for this process has
been used to set the most stringent indirect limit on
the charged Higgs mass and to constrain the magnitude
of the effective Wilson coefficient of the electromagnetic
penguin operator $|C_7^{\mathrm{eff}}|$ \cite{charged-Higgs}.
However, it cannot constrain the sign of $C_7^{\mathrm{eff}}$, 
which is essential to obtain definitive evidence of new physics 
since $C_7^{\mathrm{eff}}$ is negative in the SM 
while it can be positive in some non-SM physics models \cite{C7}.
The electroweak penguin decays $B \to X_{s} \ell^{+} \ell^{-}$ 
are promising from this point of view since the coefficients 
$C_7^{\mathrm{eff}},\ C_9^{\mathrm{eff}}$ and $C_{10}$ 
can be determined by measuring 
the dilepton invariant mass distributions and
forward-backward charge asymmetry 
of the dilepton and the $B\to X_{s}\gamma$ 
decay rate\cite{Ali-1994}.

Standard Model branching fraction predictions for
$B \to K^{(*)}\ell^{+}\ell^{-}$ decays are 
listed in Table~\ref{tab:brpred} 
[\ref{ref:Ali}--\ref{ref:Melikhov}].
Although several groups \cite{EXP} have 
searched for 
$B\to K^{(*)}\ell^+\ell^-$ decays,
no evidence has been observed. 

In this Letter, we present the 
results of a search for $B$ decays to $K^{(*)} \ell^+ \ell^-$
using data collected with the Belle detector 
at the KEKB storage ring\cite{kekb}. 
The data sample corresponds to 29.1~fb${}^{-1}$ taken 
at the $\Upsilon(4S)$ resonance and contains 
31.3 million $B\overline{B}$ pairs.

Belle is a general-purpose detector based on a 1.5~T 
superconducting solenoid magnet that surrounds the KEKB beam 
crossing point. 
Charged particle tracking is provided by a Silicon Vertex 
Detector (SVD) and a Central Drift Chamber (CDC).
Particle identification is accomplished by a combination of 
silica Aerogel \v{C}erenkov Counters (ACC),
a Time of Flight counter system (TOF)
and specific ionization measurements ($dE/dx$) in the CDC. 
A CsI(Tl) Electromagnetic Calorimeter (ECL) is located inside 
the solenoid coil.
The $\mu/K_L$ detector (KLM)
is located outside of the coil. 
A detailed description of the Belle detector can be found 
elsewhere\cite{NIM}.

In this analysis, charged tracks, except for the $K^{0}_{S} \to \pi^{+} \pi^{-}$
decay daughters, are required to have a point of 
closest approach to the interaction point 
within 0.5~cm in the $r\phi$ plane and 
5.0~cm in the $z$ direction,
where the $r\phi$ plane is the plane perpendicular to the electron-beam ($z$) direction.
Electrons are identified from the ratio of shower energy 
in the ECL to the momentum measured by the CDC, the shower shape 
of the cluster in the ECL, $dE/dx$ in the CDC 
and the light yield in the ACC.
Tracks are identified as muons based on the matching quality
and penetration depth of associated hits in the KLM.
To reduce the misidentification of hadrons as leptons, 
we require that the momentum be greater than $0.5$~GeV/$c$ 
and $1.0$~GeV/$c$ for electron and muon candidates, respectively.
Charged kaons and pions are identified by a likelihood ratio based on 
$dE/dx$ in the CDC, time-of-flight information and the ACC response.

Photons are selected from isolated showers in the ECL 
with energy greater than $50$~MeV and a shape 
that is consistent with an electromagnetic shower. 
Neutral pion candidates are reconstructed from pairs of
photons, and are required to have an invariant mass 
within $10$~MeV/$c^2$ of the nominal $\pi^{0}$ mass and
a laboratory momentum greater than 0.1 GeV/$c$.
$K^{0}_{S}$ candidates are reconstructed from oppositely 
charged tracks with a vertex displaced from 
the interaction point. 
We require the invariant mass to lie within $15$~MeV/$c^{2}$ 
of the nominal $K^{0}_{S}$ mass.

$K^{*}$ candidates are formed by combining 
a kaon and a pion: $K^{+}\pi^{-}$, $K^{0}_{S}\pi^{0}$, 
$K^{0}_{S}\pi^{+}$ or $K^{+}\pi^{0}$\cite{footnote1}.
The $K^{*}$ invariant mass is required to lie within 
$75$~MeV/$c^2$ of the nominal $K^{*}$ mass. 
For modes involving $\pi^{0}$'s, 
combinatorial backgrounds
are reduced by the further requirement
$\cos\theta_{\rm{hel}}<0.8$,
where $\theta_{\rm{hel}}$
is defined as the angle between the $K^{*}$ momentum direction and 
the kaon momentum direction in the $K^{*}$ rest frame.

$B$ candidates are reconstructed from a $K^{(*)}$ candidate 
and an oppositely charged lepton pair. 
Backgrounds from the $B \to J/\psi(\psi^{'}) K^{(*)}$
are rejected using the dilepton invariant mass veto windows;
$ -0.25 < M_{ee}     - M_{J/\psi} < 0.07~\textrm{GeV/}c^{2}$ for $J/\psi\ K^* $,
$ -0.20 < M_{ee}     - M_{J/\psi(\psi^{'})} < 0.07~\textrm{GeV/}c^{2}$ for $J/\psi\ K(\psi^{'}K^{(*)})$,
$ -0.15 < M_{\mu\mu} - M_{J/\psi}   < 0.08~\textrm{GeV/}c^{2}$ for $J/\psi\ K^{*}$ and
$ -0.10 < M_{\mu\mu} - M_{J/\psi(\psi^{'})}   < 0.08~\textrm{GeV/}c^{2}$ for $J/\psi\ K(\psi^{'}K^{(*)})$.
To suppress the background from photon conversions and $\pi^{0}$ Dalitz decays,  
we require the dielectron mass to satisfy $M_{ee} > 0.14~$GeV/$c^{2}$.

Backgrounds from continuum $q\overline{q}$ events are suppressed 
using event shape variables. 
A Fisher discriminant $\cal{F}$\cite{fd} is calculated from 
the energy flow in 9 cones along the $B$ candidate sphericity axis 
and the normalized second Fox-Wolfram moment $R_{2}$\cite{fw}. 
Furthermore, we use the $B$ meson flight 
direction $\cos\theta_{B}$ and the angle between the $B$ meson 
sphericity axis and the $z$ axis, $\cos\theta_{\rm{sph}}$. 
For the muon mode, $\cos\theta_{\rm{sph}}$ is not used 
since its distribution is nearly the same for signal and 
continuum due to detector acceptance. 
We combine $\cal{F}$, $\cos\theta_{B}$ and $\cos\theta_{\rm{sph}}$ 
into one likelihood ratio ${\cal{LR}}_{\rm{cont}}$ defined as
$
  {\cal{LR_{\rm{cont}}}} 
= {\cal{L_{\rm{sig}}}}/({\cal{L_{\rm{sig}}} + \cal{L_{\rm{cont}}}}),
$
where ${\cal{L_{\rm{sig}}}}$ and ${\cal{L_{\rm{cont}}}}$ are 
the products of the probability density functions 
for signal and continuum background, 
respectively. 

The major background from $B\overline{B}$ events is due to semileptonic $B$ decays.
The missing energy of the event, $E_{\mathrm{miss}}$, 
is used to suppress this background since we expect a large amount of missing 
energy due to the undetected neutrino.
The $B$ meson flight angle $\cos\theta_{B}$ is also used to suppress combinatorial 
background in $B\overline{B}$ events. 
We combine $E_{\mathrm{miss}}$ and $\cos\theta_{B}$ into 
the likelihood ratio ${\cal{LR}}_{B\overline{B}}$, defined similarly to ${\cal{LR_{\rm{cont}}}}$.

Finally, we calculate 
the beam-energy constrained mass $M_{\mathrm{bc}}=\sqrt{E^2_{\mathrm{beam}}-p^2_B}$ 
and the energy difference $\Delta E = E_{B}-E_{\mathrm{beam}}$ 
to select $B$ candidates, where $E_{\mathrm{beam}} = \sqrt{s}/2$
is the beam energy in the center of mass (cm) frame and
$p_B$ and $E_{B}$ are the measured momentum and energy of 
the $B$ candidate in the cm frame, respectively.
The selection criteria are tuned to maximize the expected 
significance $S/\sqrt{S+B}$ where $S$ is the signal yield 
and $B$ is the expected background in the signal box. 
$S$ and $B$ are determined from GEANT based Monte Carlo (MC) samples.
The $B\to K^{(*)}\ell^+\ell^-$ decays are generated according to 
the Greub, Ioannissian and Wyler model\cite{Greub} with 
the branching fractions predicted by Ali {\it et al.}\cite{Ali}. 
The interference between $K^{(*)}\ell^+\ell^-$ and 
$J/\psi(\psi^{'}) K^{(*)}$ 
is not considered.
The signal box is defined as $|M_{\mathrm{bc}}-M_B|<7$~MeV/$c^{2}$ 
($2.7\sigma$) for both the electron mode and the muon mode,
where $M_B$ is the nominal $B$ meson mass, 
and $-0.06<\Delta E<0.04$~GeV for the electron mode 
and $|\Delta E|<0.04$~GeV for the muon mode. 
We make selections on ${\cal{LR}}_{\mathrm{cont}}$ and ${\cal{LR}}_{B\overline{B}}$
that reject 85\% of the continuum background and 45\% of the $B\overline{B}$ background
and retain 75\% of the signal for all modes except for those with
$K^{0}_{S}\pi^{+}$ and $K^{+}\pi^{0}$ final states, where the selection 
on ${\cal{LR}}_{B\overline{B}}$ is tightened to reject 55\% of the $B\overline{B}$ background
and retain 70\% of the signal.
The overall detection efficiencies, estimated by the MC simulation, are listed in Table \ref{tab:results}.

\begin{figure}
  \begin{center}
    \epsfxsize 4.0 truein \epsfbox{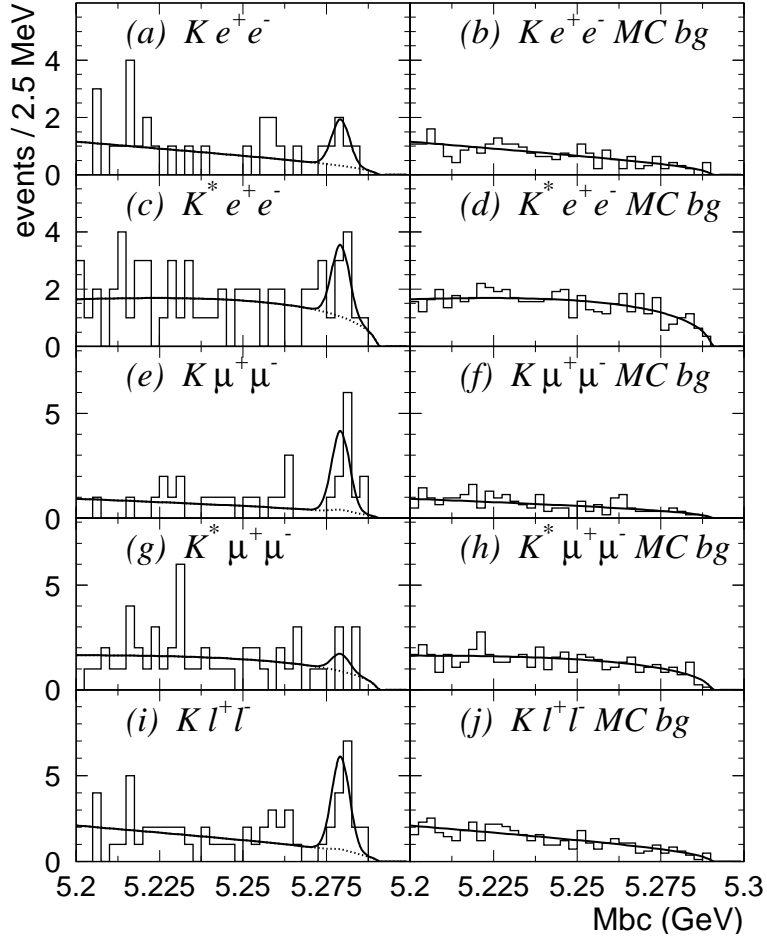}
  \end{center}
  \caption{$M_{\mathrm{bc}}$ distributions with fits for
   (a,b)~$B \to K e^{+} e^{-}$,   (c,d)~$B \to K^{*} e^{+} e^{-}$,
   (e,f)~$B \to K \mu^{+} \mu^{-}$, (g,h)~$B \to K^{*} \mu^{+} \mu^{-}$
   and (i,j)~$B \to K \ell^{+} \ell^{-}$.
    Left column is for data and right column is for MC background.
   The solid curve in the right column shows the fit results with
   the ARGUS function, which is used in the fit to the data in the left column.
   }
  \label{fig:mbcfitbgshape}
\end{figure}

To determine the signal yield, 
we perform a binned maximum-likelihood fit to 
each $M_{\mathrm{bc}}$ distribution.
The expected number of events  is 
calculated as a function of $M_{\mathrm{bc}}$,  from a Gaussian  signal distribution plus  background functions.
The mean and the width of the signal Gaussian 
are determined using observed $J/\psi K^{(*)}$ events.
A MC study shows that the width has no dependence  on the dilepton invariant mass.
The background from real leptons is parameterized by 
the ARGUS function\cite{argus}.
The shape is determined from 400~fb${}^{-1}$ MC samples,
each containing at least one oppositely charged lepton pair.
As shown in Figure~\ref{fig:mbcfitbgshape} (right column), the ARGUS function is a good representation
of the background distributions.
The MC shape is consistent with the shapes derived from the $\Delta E$ 
sideband and the $K^{(*)} e^{\pm}\mu^{\mp}$ samples in the data.
The background contribution due to misidentification of hadrons 
as muons is  parameterized
by another ARGUS function and a Gaussian.
The ARGUS function represents the combinatorial background while
the Gaussian represents the background that makes a peak
in the signal box. 
The shape and normalization of this background are 
fixed using the $B\to Kh^+h^-$ data sample ($h^\pm$ refers to hadrons).
All $K h^{+} h^{-}$ combinations are weighted by the momentum dependent
probability of misidentifying $K h^+h^-$ as $K\mu^+\mu^-$.
This study yields $0.27 \pm 0.03$ $K h^+h^-$ events in the peak region.
For electron mode, the misidentified $K h^+ h^-$ background in the peak region
is less than 0.007 events. Other backgrounds with
misidentified leptons are negligible.
The normalizations of the signal and the background from real leptons are floated in the fit. 

The fit results are shown in Figure~\ref{fig:mbcfitbgshape} (left column) and
summarized in Table \ref{tab:results}.
The statistical significance is defined as 
$\sqrt{-2\ln({\cal L}_0/{\cal L}_{\rm{max}})}$, 
where ${\cal L}_{\rm{max}}$ is the maximum likelihood 
in the $M_{\mathrm{bc}}$ fit and ${\cal L}_0$ is the likelihood 
when the signal yield is constrained to be zero.
We observe 11 $K\mu^+\mu^-$ events.
The fit to the $M_{\mathrm{bc}}$ distribution yields
 $9.5^{+3.8}_{-3.1}$ signal and $1.6\pm 0.4$ background events.
The statistical significance of this excess is 4.7.
The probability of an upward fluctuation of the background to 
11 or more events is
$5.5 \times 10^{-6}$,
which corresponds to  4.4 standard deviations for a Gaussian probability distribution.
As a test we also perform
a fit to the $\Delta E$ distribution and
find a signal yield of $8.5^{+3.7}_{-2.4}$, which
is consistent with the $M_{\mathrm{bc}}$ fit results.

\begin{figure}
\begin{center}
\epsfxsize 4.0 truein \epsfbox{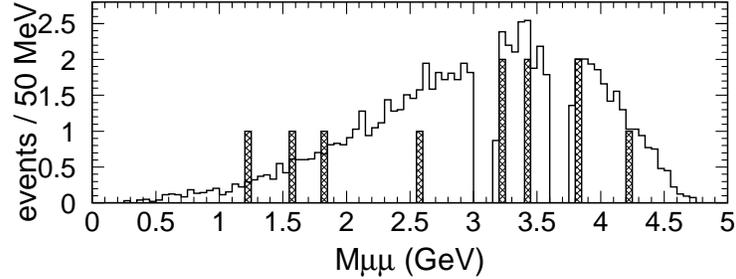}
\end{center}
\caption{
Dimuon mass distribution of $B\to K \mu^{+} \mu^{-}$ candidates.
The hatched histogram shows the data distribution while the open
histogram shows the MC signal distribution.
}
\label{fig:kmumu-mass}
\end{figure}

The kinematical properties of the $K\mu^{+} \mu^{-}$ events 
are further examined to check for potential backgrounds
that might peak in the signal area.  
The $B^{+}\to \overline{D}{}^{0}\pi^+,~\overline{D}{}^{0}\to K^{+}\pi^-$ 
decay chain is the largest expected source of $K h^+h^-$ background.
We expect $0.20 \pm 0.12$ events based on a MC simulation study.
Another possible background source is double-misidentification 
of the $B\to J/\psi K,~J/\psi\to \mu^+\mu^-$ decay chain 
where the kaon and a muon are misidentified as a muon and a kaon, 
respectively.
The $K^+\mu^-$ combinations with $K^+\pi^-$ and $\mu^+\mu^-$ 
hypotheses are examined for the candidate events, 
and show no cluster in the $D^0$ mass or $J/\psi$ mass region, 
which confirms the MC expectation.
The $B\to J/\psi X,~J/\psi\to \mu^+\mu^-$ decay chain can be 
another background source when muon pairs from 
$J/\psi\to \mu^+\mu^-$ decays evade the $J/\psi$ veto.
We expect 0.08 events using a MC sample.
The $\mu$ pair effective mass distribution is consistent with 
the MC expectation (Fig.~\ref{fig:kmumu-mass}), and 
we observe no events close to the $J/\psi$ or $\psi'$ veto region. 
To summarize, we observe no indication of a background producing a peak
in the $M_{\mathrm{bc}}$ distribution in the $K \mu^{+} \mu^{-}$ 
sample.

We consider systematic effects from the fit and the efficiency 
determination.
Uncertainty in the background function is the dominant source 
of the systematic error.
To evaluate the effect of the signal function parameters,  
the mean and the width of the Gaussian
are changed by $\pm 1 \sigma$ from the values 
determined from $J/\psi K^{(*)}$ events.
The uncertainty in the background shape is obtained by varying 
the ARGUS shape parameter by $\pm 1 \sigma$ from the value determined 
with a large MC sample. The magnitude of the variation is 
rescaled to an equivalent luminosity of 29.1~fb${}^{-1}$.
Even if the background shape is modified to maximize the background
contribution in the signal region, 
the statistical significance of the $K\mu^+\mu^-$ signal remains 
above 4.0. 
The systematic errors associated with the fit function are shown 
in the third column of Table \ref{tab:results}.
Systematic uncertainties on the tracking, charged kaon ID, 
charged pion ID, electron ID, muon ID, $K^{0}_{S}$ detection 
and $\pi^{0}$ detection efficiencies are estimated to be 2.3 to 2.5\%, 
2.1 to 2.5\%, 0.8\%, 1.8\%, 2.2\%, 8.7\% and 6.8\% 
per particle, respectively.

\vskip 3mm

%
%

In calculating the branching fraction, we assume
equal fractions of charged and neutral $B$ meson pair production 
at the $\Upsilon (4S)$. 
We combine neutral and charged $B$-meson results for 
$B\to K \mu^{+} \mu^{-}$ modes 
and obtain the branching fraction
\begin{center}
	${\cal B}(B \to K \mu^{+} \mu^{-}) = (0.99^{+0.40}_{-0.32}{}^{+0.13}_{-0.14}) \times 10^{-6}$,
\end{center}
where the first and second errors are statistical and systematic, respectively.
If we combine the $B \to K e^+e^-$ and $B \to K \mu^+\mu^-$ decay modes,
we observe $13.6^{+4.5}_{-3.8}$ signal events with a statistical significance of 5.3.
Assuming lepton universality, 
the branching fraction is determined to be
\begin{center}
	${\cal B}(B \to K \ell^{+} \ell^{-}) = (0.75^{+0.25}_{-0.21} \pm 0.09 ) \times 10^{-6}$.
\end{center}
These values are consistent with the SM predictions [\ref{ref:Ali}--\ref{ref:Melikhov}].
For the modes with significance of less than 3.0, we also set
upper limits for the branching fractions, 
employing the approach of Feldman and Cousins
\cite{fc},
as listed in Table \ref{tab:results}.
These limits are consistent with  SM predictions.
%

In summary, we have observed the electroweak penguin decay
$B \to K\ell^+\ell^-$.
The branching fractions obtained can be used to constrain contributions 
of new physics in the Wilson coefficients $C_9^{\mathrm{eff}}$ and $C_{10}$.


We wish to thank the KEKB accelerator group for the excellent
operation of the KEKB accelerator. We acknowledge support from the
Ministry of Education, Culture, Sports, Science, and Technology of Japan
and the Japan Society for the Promotion of Science; the Australian
Research
Council and the Australian Department of Industry, Science and
Resources; the Department of Science and Technology of India; the BK21 
program of the Ministry of Education of Korea, the Basic Science program 
of the Korea Research Foundation, and the Center for High Energy 
Physics sponsored by the KOSEF; the Polish
State Committee for Scientific Research under contract No.2P03B 17017;
the Ministry of Science and Technology of Russian Federation; the
National Science Council and the Ministry of Education of Taiwan;
and the U.S. Department of Energy.

\newpage


\begin{table}[b]
  \caption{Branching fractions for $B\to K^{(*)}\ell^+\ell^-$
   predicted in the framework of the Standard Model.}
  \begin{center}
    \begin{tabular}{llll}
       & \multicolumn{3}{c}{Predicted branching fraction $[\times 10^{-6}]$}
         \\ \cline{2-4}
       \raisebox{1.5ex}[0pt]{Mode} & Ali {\it et al.}\cite{Ali}   
        & Greub {\it et al.}\cite{Greub}& Melikhov {\it et al.}\cite{Melikhov}
        \\ \hline
$K \ell^{+}\ell^{-}$ & $0.57^{+0.16}_{-0.10}$ & $0.33\pm 0.07$ & $0.42\pm 0.09$\\ 
$K^{*}e^{+}e^{-}$ & $2.3^{+0.7}_{-0.4}$ & $1.4\pm 0.3$ & $1.4\pm 0.5$ \\ 
$K^{*}\mu^{+}\mu^{-}$ & $1.9^{+0.5}_{-0.3}$ & $1.0\pm 0.2$ & $1.0\pm 0.4$\\  
    \end{tabular}
  \end{center}
  \label{tab:brpred}
\end{table}

\mediumtext

\begin{table}[t]
 \caption{Summary of the fit results and branching fractions. 
  Number of events observed in the signal box, number of signal and 
  background events estimated from the $M_{\mathrm{bc}}$ fit,
  detection efficiency of each mode, branching fraction obtained, 
  90\% confidence level upper limit of the branching fraction
  and the statistical significance of the signal. 
  The first error in the signal yield and branching fraction is 
  statistical and the second one is systematic.
  The error in the efficiency includes MC statistics and systematic error. 
  The error in the background is statistical only.
}
 \begin{center}
 \begin{tabular}{lccccccc} 
 \raisebox{-1ex}[0pt]{Mode} & Observed & Signal 
   & Back- & Efficiency & ${\cal{B}}$ 
   & U.L. & Stat. \\
 & events & yield & ground & [\%] & [$\times10^{-6}$] & [$\times10^{-6}$] & signif.\\ \hline
$K^0 e^+ e^-$       & 1 & $0.5^{+1.4}_{-0.5}{}^{+0.4}_{-0.5}$  & $0.3 \pm 0.2$ & $5.5\pm 0.6$  & -                                        & 2.7 & -    \\ 
$K^+ e^+ e^-$       & 5 & $3.5^{+2.5}_{-1.8}{}^{+0.5}_{-0.7}$  & $1.5 \pm 0.4$ & $21.6\pm 2.0$ & $0.51^{+0.37}_{-0.27}{}^{+0.09}_{-0.11}$ & 1.4 & 2.4  \\ 
$Ke^+ e^-   $       & 6 & $4.1^{+2.7}_{-2.1}{}^{+0.6}_{-0.8}$  & $1.7 \pm 0.4$ & $13.6\pm 1.3$ & $0.48^{+0.32}_{-0.24}{}^{+0.09}_{-0.11}$ & 1.3 & 2.5  \\ \hline
$K^{*0}e^+ e^-$     & 9 & $4.0^{+2.9}_{-2.2}{}^{+1.0}_{-1.1}$  & $3.9 \pm 0.7$ & $6.6\pm 0.7$  & -                                        & 6.4 & -    \\
$K^{*+}e^+ e^-$     & 4 & $2.5^{+2.3}_{-1.6}{}^{+0.3}_{-0.4}$  & $1.5 \pm 0.5$ & $3.1\pm 0.4$  & -                                        & 8.9 & -    \\ 
$K^* e^+ e^- $      &13 & $6.3^{+3.7}_{-3.0}{}^{+1.0}_{-1.1}$  & $5.7 \pm 0.9$ & $4.8\pm 0.5$  & $2.08^{+1.23}_{-1.00}{}^{+0.35}_{-0.37}$ & 5.6 & 2.5  \\ \hline\hline
$K^0\mu^+\mu^-$     & 2 & $1.9^{+1.8}_{-1.1}{}^{+0.0}_{-0.1}$  & $0.1 \pm 0.1$ & $6.5\pm 0.7$  & $0.94^{+0.88}_{-0.54}{}^{+0.11}_{-0.12}$ & 3.3 & 2.8  \\
$K^+\mu^+\mu^-$     & 9 & $7.3^{+3.4}_{-2.7}{}^{+0.9}_{-1.0}$  & $1.8 \pm 0.4$ & $23.9\pm 2.2$ & $0.98^{+0.46}_{-0.36}{}^{+0.15}_{-0.16}$ & -   & 3.9  \\ 
$K\mu^+\mu^-  $     &11 & $9.5^{+3.8}_{-3.1}{}^{+0.8}_{-1.0}$  & $1.6 \pm 0.4$ & $15.2\pm 1.4$ & $0.99^{+0.40}_{-0.32}{}^{+0.13}_{-0.14}$ & -   & 4.7  \\ \hline
$K^{*0}\mu^+\mu^-$  & 6 & $3.2^{+2.6}_{-1.9}{}^{+0.6}_{-0.7}$  & $2.2 \pm 0.5$ & $8.3\pm 0.9$  & -                                        & 4.2 & -  \\
$K^{*+}\mu^+\mu^- $ & 2 & $0.0^{+0.7}_{-0.0}{}^{+0.0}_{-0.0}$  & $2.7 \pm 0.6$ & $3.5\pm 0.4$  & -                                        & 3.9 & -    \\ 
$K^*\mu^+\mu^-    $ & 8 & $2.1^{+2.9}_{-2.1}{}^{+0.9}_{-1.0}$  & $4.9 \pm 0.8$ & $5.9\pm 0.7$  & -                                        & 3.1 & -    \\ \hline\hline
$K \ell^+\ell^-   $ &17 & $13.6^{+4.5}_{-3.8}{}^{+0.9}_{-1.1}$ & $3.3 \pm 0.5$ & $14.4\pm 1.4$ & $0.75^{+0.25}_{-0.21} \pm 0.09$          & -   & 5.3    \\ 
  \end{tabular}
\end{center}

\label{tab:results}
\end{table}

\begin{references}
\bibitem{footnote1}{
Charge conjugate modes are implied throughout this Letter.
}
\bibitem{argus}{
	H.~Albrecht {\it et al.} (ARGUS Collaboration), Phys. Lett. B {\bf 241}, 278 (1990);~%
The ARGUS function is presented as
$ax\sqrt{1-(x/E_{\rm{beam}})^{2}} \exp(b(1-(x/E_{\rm{beam}})^{2}))$, 
where $a$ and $b$ are constants that are determined from the data.
}
\end{references}
\end{document}